\documentclass[12pt,preprint]{aastex}  

\usepackage{epsfig}
\usepackage{natbib}

\shorttitle{Planet Migration with Low Viscosity}

\shortauthors{}

\begin{document}

\title{TYPE I PLANET MIGRATION IN NEARLY LAMINAR DISKS -- LONG TERM
  BEHAVIOR}

\author{C.~Yu\altaffilmark{1,2}, H.~Li\altaffilmark{2},
  S.~Li\altaffilmark{2}, S.~H. Lubow\altaffilmark{3},
 D.N.C. Lin\altaffilmark{4}}
\altaffiltext{1}{National Astronomical Observatories/Yunnan Astronomical Observatory,
Chinese Academy of Sciences, Kunming, 650011}
\altaffiltext{2}{Los Alamos National Laboratory, Los Alamos, NM 87545;
  {\tt congyu@lanl.gov, hli@lanl.gov; sli@lanl.gov}}
\altaffiltext{3}{Space Telescope Science
  Institute, 3700 San Martin Drive, Baltimore, MD 21218; {\tt
    lubow@stsci.edu}}
\altaffiltext{4}{UCO/Lick Observatory, University of California, Santa
  Cruz, CA 95064; {\tt
    lin@ucolick.org}}


\begin{abstract}
 
  We carry out 2-D high resolution numerical simulations of type I planet migration with different 
  disk viscosities. 
  We find that the planet migration is strongly dependent on disk viscosities.
  Two kinds of density wave damping mechanisms 
  are discussed.
  Accordingly, the angular momentum transport can be either
  viscosity dominated or shock dominated, depending on the disk viscosities.
  The long term migration behavior is different as well.
  Influences of the Rossby vortex instability 
  on planet migration are also discussed. 
  In addition, we investigate very weak shock generation in
  inviscid disks by small mass planets and compare the results with prior analytic results. 
 

\end{abstract}

\keywords{accretion, accretion disks --- hydrodynamics --- methods: numerical ---
planetary systems: protoplanetary disks}


\section{INTRODUCTION}
The discovery of close orbiting extrasolar planets led to extensive
studies of disk planet interactions and the forms of migration that can explain
their location. Early theoretical work established the so-called type I and 
type II migration regimes for low mass embedded planets and high mass
gap forming planets
(Goldreich \& Tremaine 1980; Lin \& Papaloizou 1986; Ward 1997), respectively. 

Although it is suggested that migration is necessary to account for the observed
distribution of planets (Ida \& Lin 2008), 
the problem is that
analytic theories and numerical simulations have shown that migration
timescales of type I are quite short (Tanaka et al. 2002) so that the
planet tends to migrate to its central star before it has time to
become massive enough to open a gap in the disk.
This problem thus becomes a competition between two timescales: type I migration
and core accretion for planet mass growth (Pollack et al. 1996; Hubickyj et al. 2005).
Several mechanisms have been suggested to address this challenging
problem, which include thermal effects of the disk (Jang-Condell 
\& Sasselov 2004), radial opacity jump (Menou \& Goodman 
2004), magnetic turbulent fluctuations 
(Nelson \& Papaloizou 2004) and effects of co-orbital material 
(Masset et al. 2006). 
Non-isothermal slowing down of type I migration is studied by
Paardekooper \& Mellema (2006), 
Baruteau \& Masset (2008), and Kley et al. (2009).

Recently, Li et al. (2009; hereafter Paper I) found that the low mass
planet  migration
can have a strong dependence on the disk viscosity.  They found that
the type I migration is halted in disks of sufficiently low
viscosity. This is caused by a density feedback effect which results
in a mass redistribution around the planet. The simulations confirm the
existence of  a critical mass ($M_{cr} \sim 10 M_{\oplus}$) beyond
which migration is halted in nearly laminar disks. The critical masses
are in good agreement with the analytic model of Rafikov (2002).

This paper is a follow-up study to Paper I. By performing a
series of high resolution, 2-D hydrodynamic simulations, we present a more detailed
analysis on the density feedback effect, and describe the long term
($> 10^4$ orbits) behavior of migration. 
The paper is organized as follows. In \S 2 we give a brief description of our simulations.
In \S 3 we discuss the density wave damping mechanism for different
disk viscosities and the consequent long term migration behavior,
including the density feedback and the Rossby vortex instability (RVI).
Possible 3-D effects are discussed in \S 4. 
Summary and discussions are given in \S 5. A study on the shock
excitation in inviscid disks is given in the Appendix. 

\section{Simulations}

The 2-D hydrodynamic simulation set-up and the numerical methods we
used here are the same as that in Paper I (more details on the code are given in
Li \& Li 2009). We choose an initial surface density profile normalized to
the minimum mass solar nebulae model (Hayashi 1981)
as $\Sigma(r) = 152 f (r/5 AU)^{-3/2} \rm{gm \ cm}^{-2}$, where $f=1$
in this paper. (The migration dependence on $f$ in the low viscosity
limit has been explored in Paper I so we will not vary $f$ in
this study.)  
The disk is assumed to be isothermal throughout the simulated region,
having a constant sound speed $c_s$.
The dimensionless disk thickness $c_s/v_{\phi}(r=r_p)$ is set as $0.035$,
where $v_{\phi}$ is the Keplerian velocity at the initial planet location $r_p$. 
(Simulations with higher $c_s$ were given in Paper I.)  
The dimensionless kinematic viscosity $\nu$ (normalized
by $\Omega^2 r$ at the planet's initial orbital radius) is taken 
as a spatial constant and ranges from $0$ to $10^{-6}$. This
corresponds to an effective viscous $\alpha = 1.5\nu/h^2 = 1.2\times 10^3 \nu$. 
For most runs, we have chosen the planet mass to be $10 M_{\oplus}$. The planet's Hill
radius is $r_H =0.0215 r_p$, which is $\sim 0.6 h$. A pseudo-3D
softening is used (Li et al. 2005). Fully 2-D disk-self
gravity is included (Li, Buoni, \& Li 2009). The disk is simulated with
$0.4 \leq r \leq 2$. Runs are made typically
using a radial and azimuthal grid of $(n_r\times n_\phi) = 800\times
3200$, though we have used higher resolution to ensure convergence on
some runs. Simulations typically last more than ten thousand orbits so
that we can study the long term behavior of migration. 

\section{Results}

\begin{figure}
\begin{center}
\epsfig{file=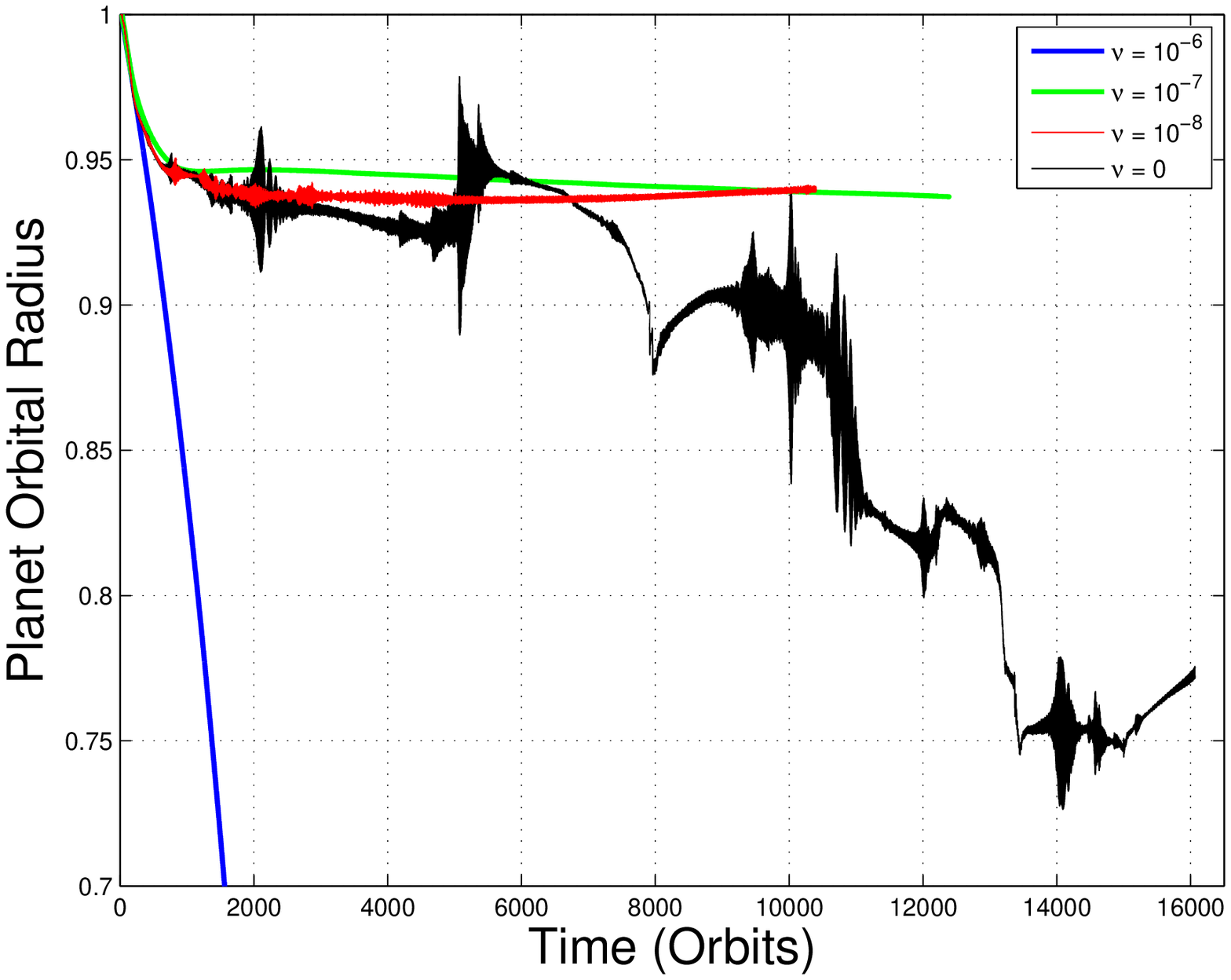,height=3in,width=5in,angle=0}
\end{center}
\caption{The orbital radius evolution of a $10 M_{\oplus}$ planet
  migrating in disks with
  different viscosities $\nu = 0, 10^{-8}, 10^{-7},$ and
  $10^{-6}$. The normalized disk sound speed is $c_s=0.035$. The
  effective $\alpha$ due to the viscosity is $
  \approx 0,  10^{-5}, 10^{-4},$ and $10^{-3}$, respectively. 
        }
\label{09v_runs}
\end{figure}

Figure \ref{09v_runs} shows the orbital radius evolution of a $10
M_{\oplus}$ protoplanet in a 2-D laminar disk with different disk viscosities. 
When the disk viscosity is relatively large ($\nu = 10^{-6}$ or
$\alpha = 1.2\times 10^{-3}$), the migration rate
agrees well with the theoretical results given by Tanaka et al. (2002)
for type I migration. When the disk viscosity is low, 
the migration behavior differs markedly from the usual type I
migration (see also Fig. 1 in Paper I). Such slowing down behavior 
was explained in terms of the density feedback effect (Ward 1997;
Rafikov 2002) in Paper I. Here, we have extended the evolution to be
about ten  times longer
($> 10^4$ orbits) than those in Paper I. But before we discuss the
long term behavior in detail, we present some additional analysis of the
density feedback effect first. 

\subsection{Density Wave Damping Mechanism for Different Viscosities}

The critical physics issue in deciding the density feedback effect is
the density wave damping mechanism. Where and how the density waves
generated by the protoplanet damp will contribute critically to the
torque on the planet. Furthermore, such damping process will modify
the density distribution around the planet, which directly affects the
torque as well. This effect was partially analyzed in Paper I. In
principle, the density wave can damp both due to disk viscosity (a
viscous process) and by shocks (a nonlinear process). The relative
importance of these processes will naturally depend on the disk
viscosity. 

To quantify the damping process, we have evaluated the Reynolds
stress and viscous stress. An effective, azimuthally averaged $\alpha$
based on the Reynolds stress can be defined as:
\begin{equation}
\alpha_{Rey}
= \ \left\langle\frac{\Sigma v_r  
\delta v_{\phi}}{P}\right\rangle~~,
\end{equation}
where $\langle ... \rangle$ indicates the azimuthal average, $\delta v_{\phi} =
v_{\phi} -  \langle v_{\phi}\rangle$, $\Sigma$ and $P$ are disk
surface density and pressure, respectively. This method was previously
discussed in Balbus \& Hawley (1998) and Li et al. (2001). 
Similarly, the azimuthally
averaged $\alpha$ based on the viscous stress can be defined as:
\begin{equation}
\alpha_{vis}
= \ \left\langle\frac{\nu \Sigma r
    \frac{d\Omega}{dr}} {P}\right\rangle~~,
\end{equation}
and this quantity scales as $r^{-3/2}$. 

\begin{figure}
\begin{center}
\epsfig{file=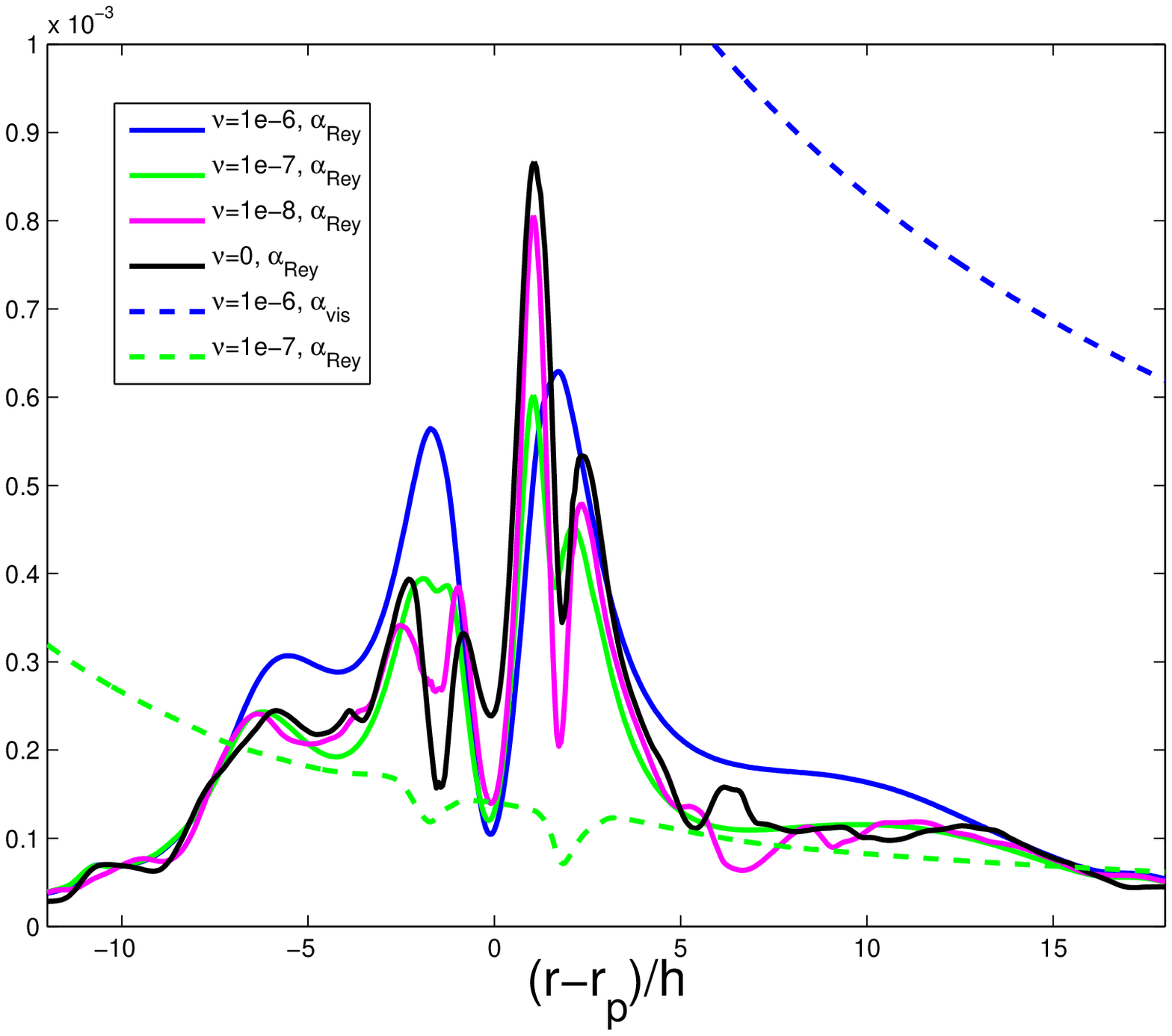,height=3in,width=5in,angle=0}
\end{center}
\caption{Comparison of the azimuthal averaged Reynold stress
         $\alpha_{Rey}$ (solid lines)
         and the viscous stress $\alpha_{vis}$ (dashed lines). Note
         that $\alpha_{vis}$  roughly scales as
         $r^{-3/2}$. For $\nu = 10^{-8}$ and $\nu=0$, $\alpha_{vis}$ is not shown. 
         For $\nu = 10^{-6}$, the viscous
         stress is larger than the Reynolds  stress.
         For lower values of $\nu = 10^{-7}, 10^{-8}$ and $0$, the viscous stress is
         smaller than the Reynolds  stress. 
         The angular momentum transport is shock dominated
         when the viscosity $\nu \le 10^{-7}$ or $\alpha \le 10^{-4}$. 
        }
\label{reynoldsviscous}
\end{figure}

In Figure \ref{reynoldsviscous}, we present both
$\alpha_{Rey}$ and $\alpha_{vis}$ as a
function of disk radius. The results are based on the runs at 400, 1000, 760 and 700 orbits 
for $\nu = 10^{-6}, 10^{-7}, 10^{-8}$ and $0$, respectively. 
These times are chosen so that the planet is at roughly the same
orbital radius in all the runs. 
For $\nu=10^{-6}$, the viscous transport is
much bigger than the Reynolds transport (by more than a factor of 2
around the planet).  
For $\nu=10^{-7}$, $ \alpha_{vis} \approx 1.2\times 10^{-4}$, which is
smaller than $\alpha_{Rey}$ around the planet. This means that the
dominant wave damping mechanism around the planet changes from being
viscous damping to being shock-dominated damping when the
disk viscosity changes from $10^{-6}$ to $10^{-7}$. For even smaller
disk viscosity, shocks dominate the wave damping. The peaks of
$\alpha_{Rey}$ are approximately $h (=1.6 r_H)$ away from the
planet, consistent with the excitation of shocks.  Note that as disk
viscosity changes, the shock strength and structure will be changed
somewhat. This could account for the changes in $\alpha_{Rey}$ for
$\nu \leq 10^{-7}$.

\begin{figure}
\begin{center}
\epsfig{file=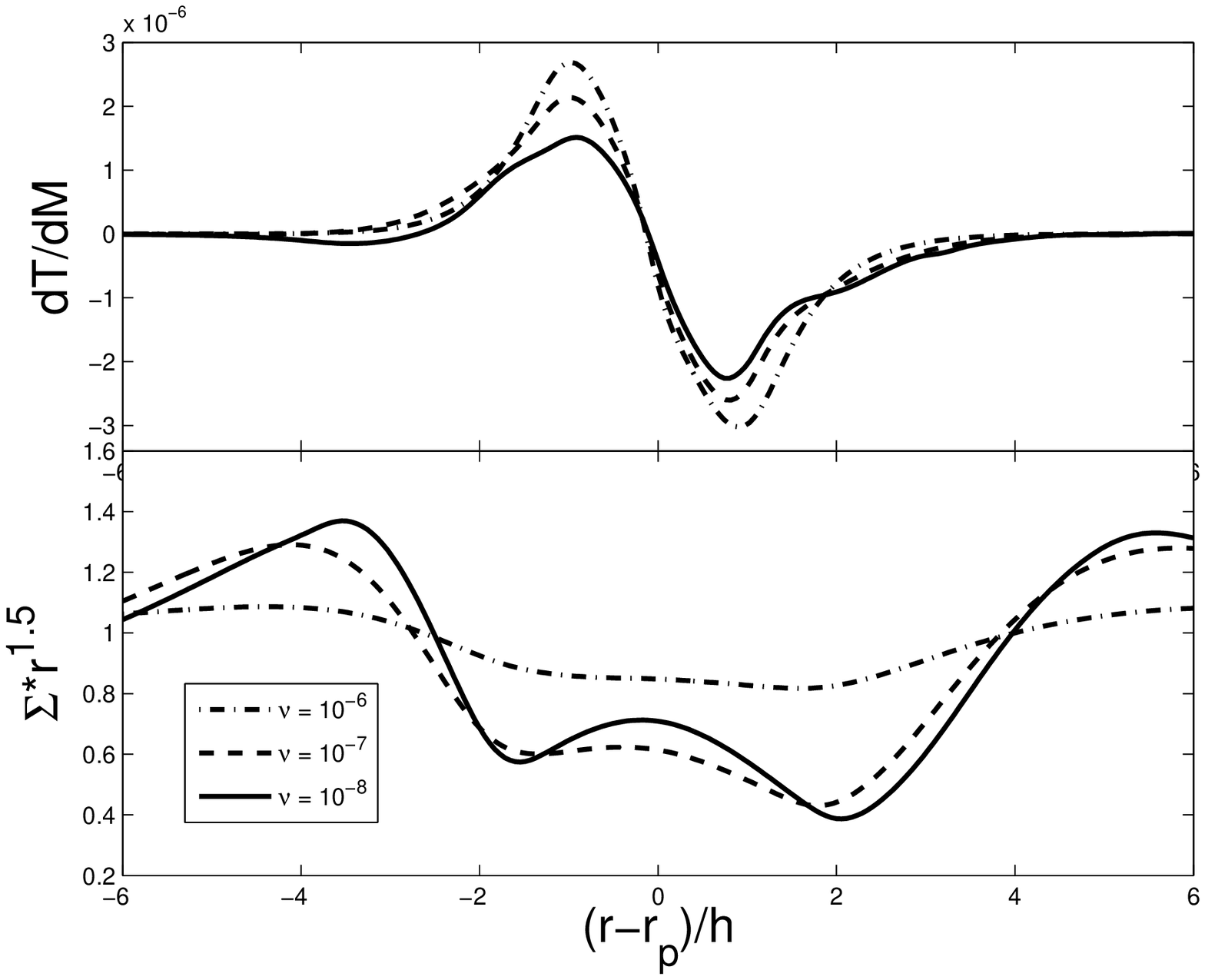,height=3in,width=5in,angle=0}
\end{center}
\caption{({\it top}) The radial profile of torque density $dT/dM$ for three
  different viscosity runs at $t = 400, 1000,$ and $760$ for 
  $\nu = 10^{-6}, 10^{-7}$, and $10^{-8}$ respectively. The planet is
  at $r_p= 0.945$ for all three cases. 
  ({\it bottom}) The azimuthally averaged surface density profiles for
  three cases. The imbalance in surface density between the inner and
  outer Lindblad regions becomes more pronounced in the lower
  viscosity runs.    
        }
\label{dTdMcmp09v1v2v3}
\end{figure}

The wave damping by shocks causes the density profiles at the shocks
to be
significantly modified. To confirm this effect further, we analyze the
torque density profiles by examining $dT/dM(r)$ where $T$ is the
torque on the planet by disk material and $M$ is the mass within each
radial ring. We choose three runs with $\nu = 10^{-6}, 10^{-7},
10^{-8}$ and pick the planet radial location at $r_p = 0.945$ to
compare (this corresponds to $t=400, 1000,$ and $760$ orbits for these
runs respectively). From Figure \ref{09v_runs}, their migration trend at this
location is quite different (i.e., the total torque on the planet is
very different). The $dT/dM(r)$ profiles are given in the top panel of Figure
\ref{dTdMcmp09v1v2v3}. For different viscosities, the difference in $dT/dM$ is
not large, within a factor of 2. But the torque amplitude on the planet in
the $\nu=10^{-6}$ case is about a factor of 100 larger than that of the two
cases for $\nu = 10^{-7}$ and $10^{-8}$. This shows that the
difference should be caused by the density variations. The bottom
panel of Figure \ref{dTdMcmp09v1v2v3}
shows the radial disk density profile around the planet. The density
imbalance between the inner Lindblad and outer Lindblad regions are
much stronger for the lower viscosity cases than the case for $\nu = 10^{-6}$. 

In the usual picture of wave damping, as the viscosity decreases, the
density waves are expected to propagate farther away (Takeuchi et al. 1996). As a
result, the peak positions of $dT/dM$ are expected to be farther
away from the planet. We did not find such behavior in the
simulation results because the shock dissipation dominates the wave
damping when the viscosity is sufficiently small. This also indicates
that, for these choices of planet mass and disk sound speed, shocks are always
produced. 

We want to emphasize that, even though the effective ``viscosity''
caused by shocks is not high (see Figure \ref{reynoldsviscous}), it is
the density imbalance due to the angular momentum transport by the
shocks that causes a big change in the total torque on the planet. For
planet's mass above the critical values described in Rafikov (2002)
and Paper I, the above analysis indicates that there exists a critical
disk viscosity, below which the density wave damping will be dominated
by the shock dissipation and the density feedback effect can slow down
(or halt) the migration. For planet mass less than the critical
values, however, even when the disk viscosity is low enough so that the shock
dissipation is dominant, the density imbalance caused by the shock
dissipation is too weak or taking too long to be able to change the  
migration behavior. 

\subsection{Long Term Evolution}

The long term evolution ($\sim 10^4$ orbits) for different disk viscosity
(Figure \ref{09v_runs}) is complicated. For three low viscosity cases,
$\nu = 10^{-7}, 10^{-8}$, and $0$, the migration is significantly
slowed down or even reversed, but the detailed behavior is different.
(Note that for $\nu=0$, there is still some low level of numerical
viscosity, which we estimate to be roughly equivalent to $\nu < 10^{-9}$.) 
We now discuss each case in detail.

\subsubsection{The $\nu = 10^{-7}$ case}

\begin{figure}
\begin{center}
\epsfig{file=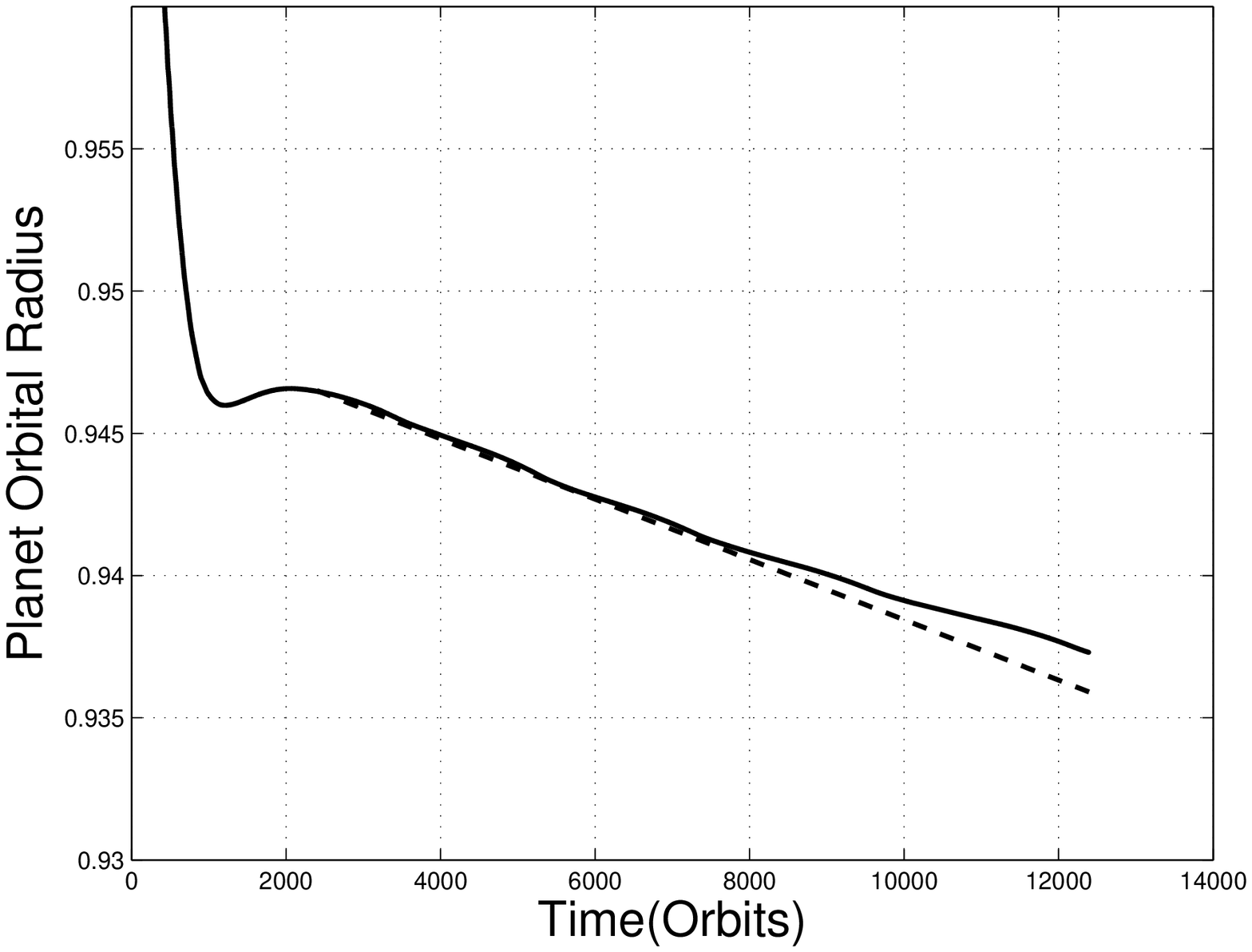,height=3in,width=5in,angle=0}
\end{center}
\caption{        Solid line is the simulation result for $\nu = 10^{-7}$.
        Dashed line is the expected viscous drift rate with $\nu = 10^{-7}$.
        The planet migration is consistent with the viscous time scale 
        after about 2500 orbits.
        }
\label{vistimescale}
\end{figure}

\begin{figure}
\begin{center}
\epsfig{file=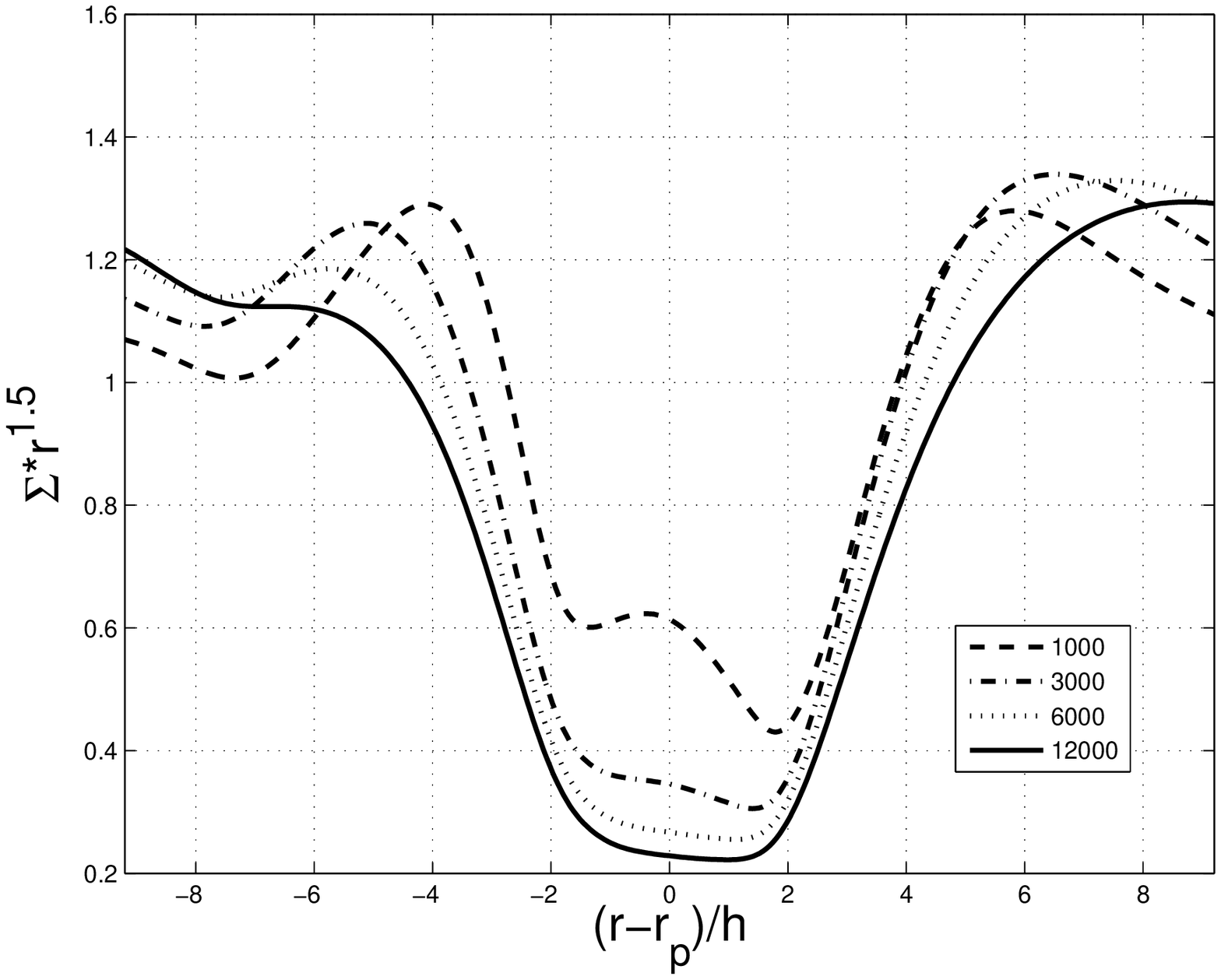,height=3in,width=5in,angle=0}
\end{center}
\caption{Surface density (azimuthally averaged)
        evolution for the $\nu =10^{-7}$ case. A smooth density
        ``depression''  forms around
        the planet and gradually widens. 
        }
\label{denevol1e-7}
\end{figure}

For $\nu = 10^{-7}$, Figure \ref{09v_runs} shows that the planet
has a steady migration rate. 
Figure \ref{vistimescale} shows the comparison between the simulation
and the viscous drift rate calculated using ${\dot r} = 3\nu/2 r_p$. 
It looks like that the density feedback effects take the planet
migration into a ``viscous'' limit, where the migration is consistent
with being on the viscous time scale after about 2500 orbits. The
corresponding surface
density profile evolution is shown in Figure \ref{denevol1e-7}. It can
be verified
that the density distribution remains smooth and evolves on the
viscous timescale as well. This implies that the shock damping of the
density waves causes the planet and surrounding disk material to
migrate with approximately the same timescale. This situation is similar to
the previous type II migration study where a gap has formed in the
disk. Upon more detailed analysis, however, the accretion rate throughout
the disk is not quite a constant. This suggests that the steady migration
observed so far could change if we follow it to even longer timescales.


Figure \ref{denevol1e-7} shows that a wide density ``depression'' (not
quite a gap) is forming. Given the wide gap, one might have expected
the excitation of the secondary instability such as the Rossby vortex instability 
(RVI), but this instability is suppressed in this case by the disk viscosity. 

\subsubsection{The $\nu = 10^{-8}$ case}

\begin{figure}
\begin{center}
\epsfig{file=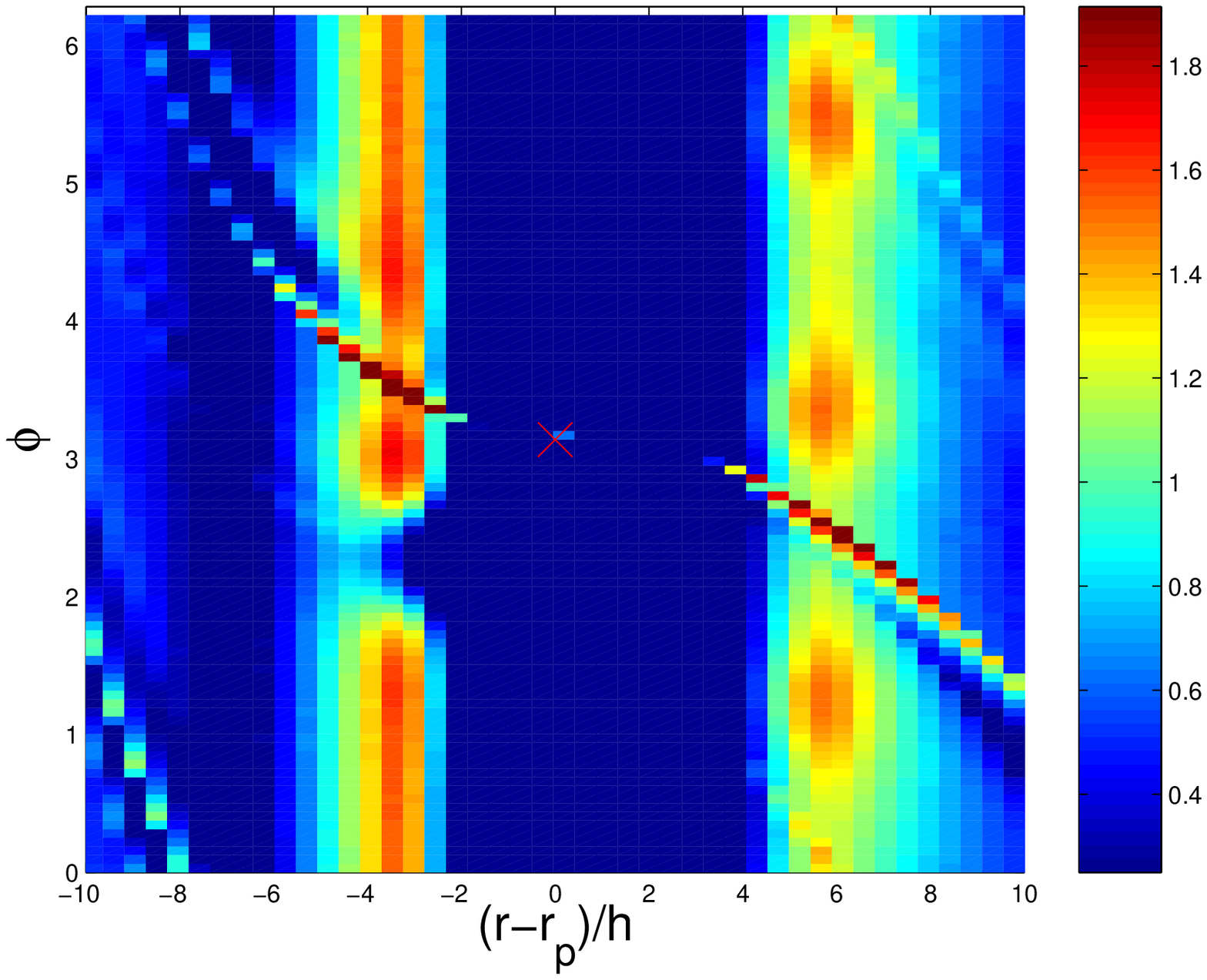,height=3in,width=5in,angle=0}
\end{center}
\caption{Vortices are excited
        for $\nu = 10^{-8}$ due to the Rossby vortex instability. 
        The 2-D disk surface density, $\Sigma\cdot r^{3/2}$, is shown.
        The time is 800 orbits.
        }
\label{den1e-8}
\end{figure}

For $\nu = 10^{-8}$, the planet's migration is essentially halted and
gradually going in the reverse direction at late stages. Figure
\ref{den1e-8} shows the density distribution at $t=800$ orbits. The
density  ``depression'' is steeper than what was seen in the $\nu =
10^{-7}$ case. For such a low viscosity, the RVI is also excited at a low
level. Figure \ref{den1e-8} shows that  
the azimuthal density variation is more pronounced in the low
azimuthal wave number $m$. This is because, during the nonlinear stage
of RVI, vortices will merge (Li et al. 2005), and one is often left
with only large scale variations in disk surface density. (More
detailed discussions on RVI will be given below.) 

\begin{figure}
\begin{center}
\epsfig{file=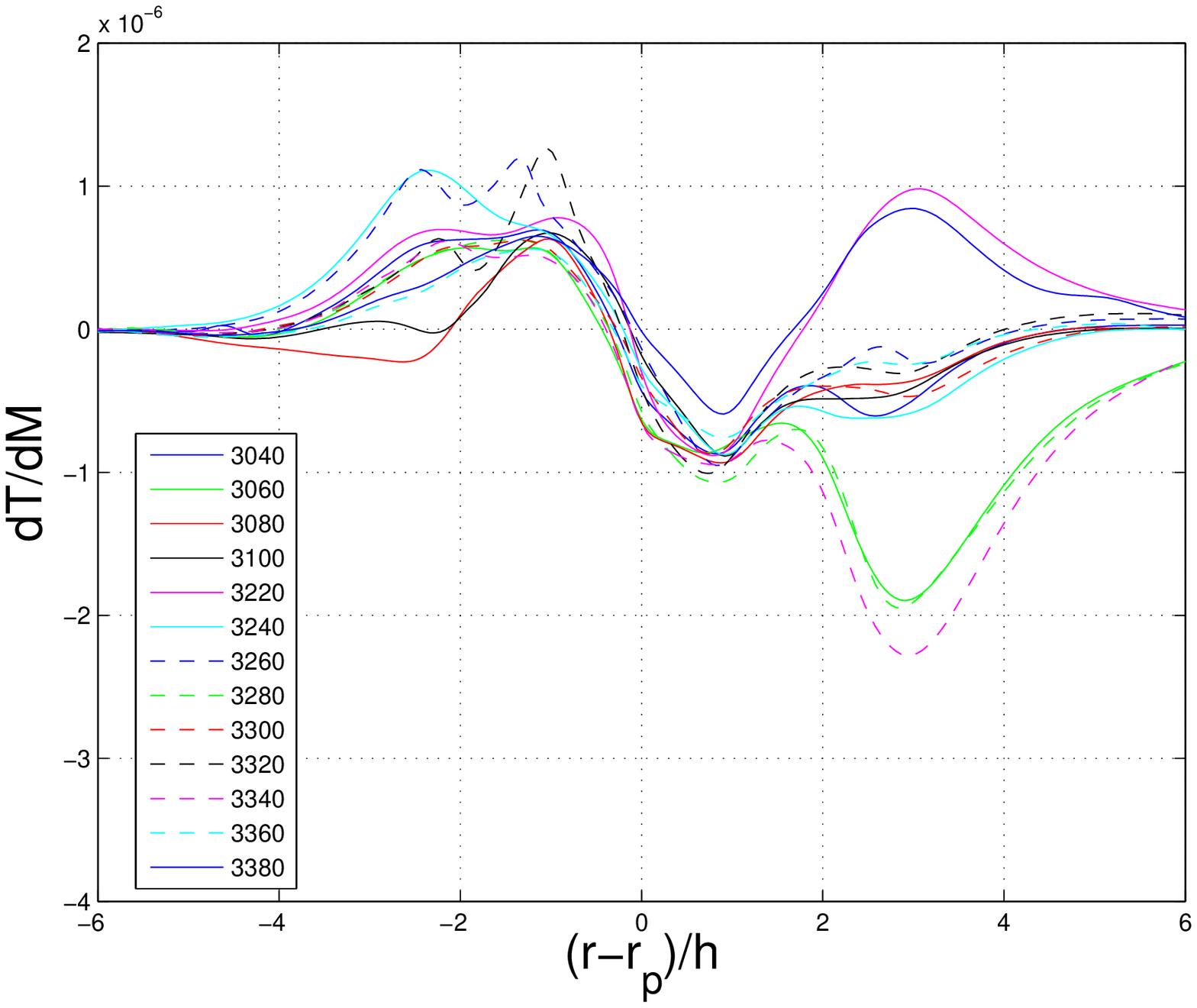,height=3in,width=5in,angle=0}
\end{center}
\caption{
Variations in $dT/dM$ as RVI is (mildly) influencing the planet
  migration during $t \sim 3000 - 3400$ orbits. 
        The peaks around $\pm h$ correspond to the shock damping. 
        The drastic variations around $\pm (2-3) h$
        are caused by RVI.
        }
\label{dTdmcmp09v3}
\end{figure}

Because of the RVI, the behavior of $dT/dM$ becomes more
complicated. This is shown in 
Figure \ref{dTdmcmp09v3} where we have plotted the evolution of
$dT/dM$ over a period between $\sim 3000 - 3400$ orbits. This
coincides with a period when the RVI is mildly excited (see Figure \ref{09v_runs}).  
The peaks around $\pm h$ in $dT/dM$ are still consistent with the shock damping.
The drastic changes around $\pm (2-3) h$ are due to the 
azimuthal asymmetries in surface density caused by RVI, which give
rise to the sign change in $dT/dM$. 
We have confirmed that the Lindblad resonance positions for these low
$m$ modes are coincident with the positions where $dT/dM$ change
dramatically around $r-r_p = \pm 2.5 h$, as shown in Figure \ref{dTdmcmp09v3}.

\begin{figure}
\begin{center}
\epsfig{file=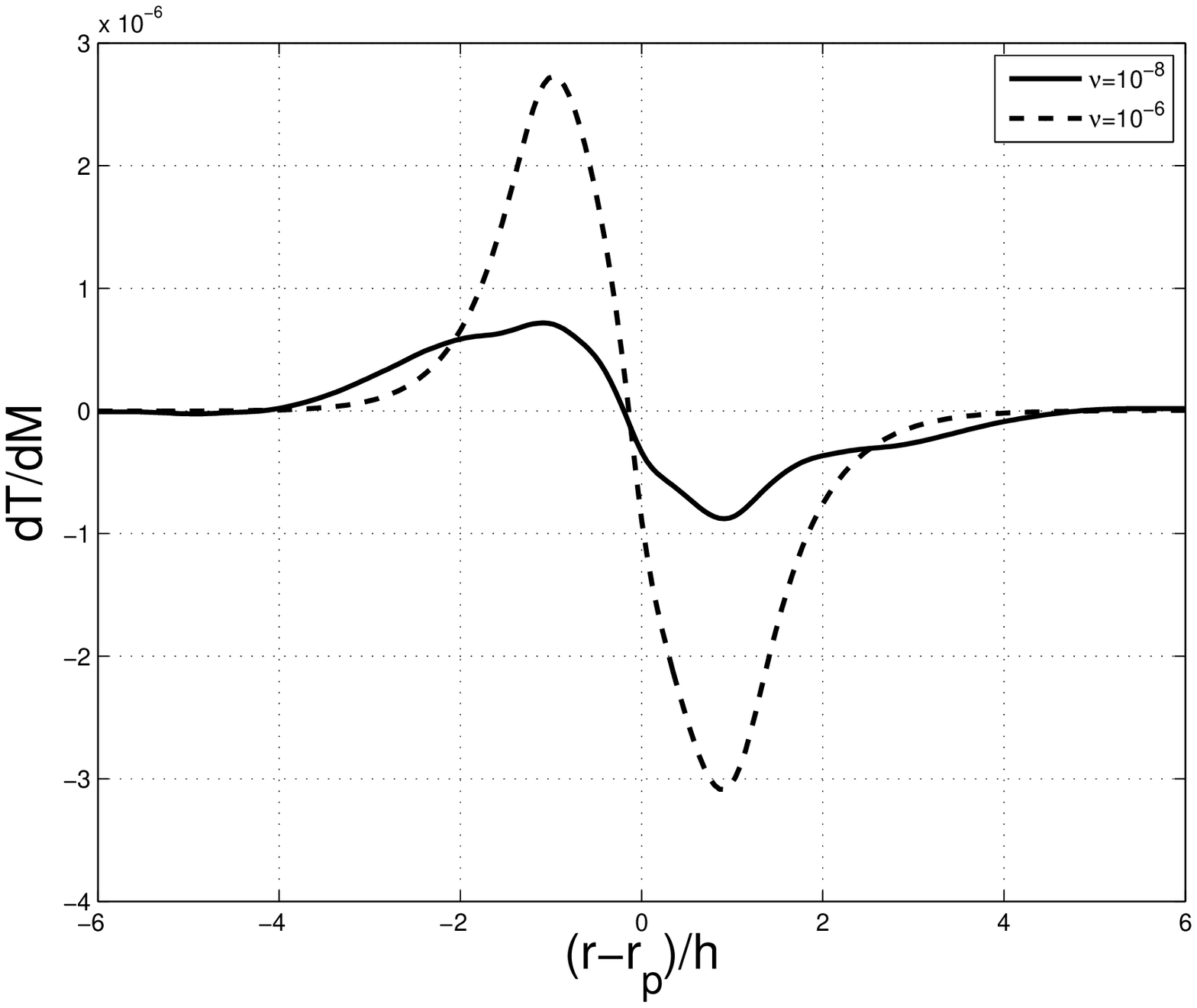,height=3in,width=5in,angle=0}
\end{center}
\caption{
        Comparison of $dT/dM$ profiles for two cases with $\nu =
        10^{-6}$  and $\nu = 10^{-8}$ when the planet is at $r_p = 0.939$.
        The solid curve is obtained by averaging over hundreds of
        orbits. 
        }
\label{dTdMcmp09v1v3}
\end{figure}

When averaged over a few hundred orbits, however, the changes in
$dT/dM$ cancel out as evident in Figure \ref{dTdMcmp09v1v3}. The
averaged profile, when compared
with that from the $\nu =10^{-6}$ case, shows that RVI causes the
torque contribution to extend to a larger radial extent (the tails
between $\pm (2.5-4) h$), though the peak amplitudes of $dT/dM$ at $\pm
h$ are smaller by about a factor of 4. This implies that the excitation
of RVI has a minor impact on the overall migration in this case. 

It is not clear why the migration is slowly going outward, nor whether
this trend will continue at much longer times than what was
simulated. This is a regime where both the density feedback effects by shock
dissipation and the influence by mild RVI are playing some roles in planet
migration. Though it seems reasonable to expect that the planet
migration is significantly slowed down when compared to the usual type
I rate, it is difficult to get a definite answer.

\subsubsection{The $\nu = 0$ case}

\begin{figure}
\begin{center}
\epsfig{file=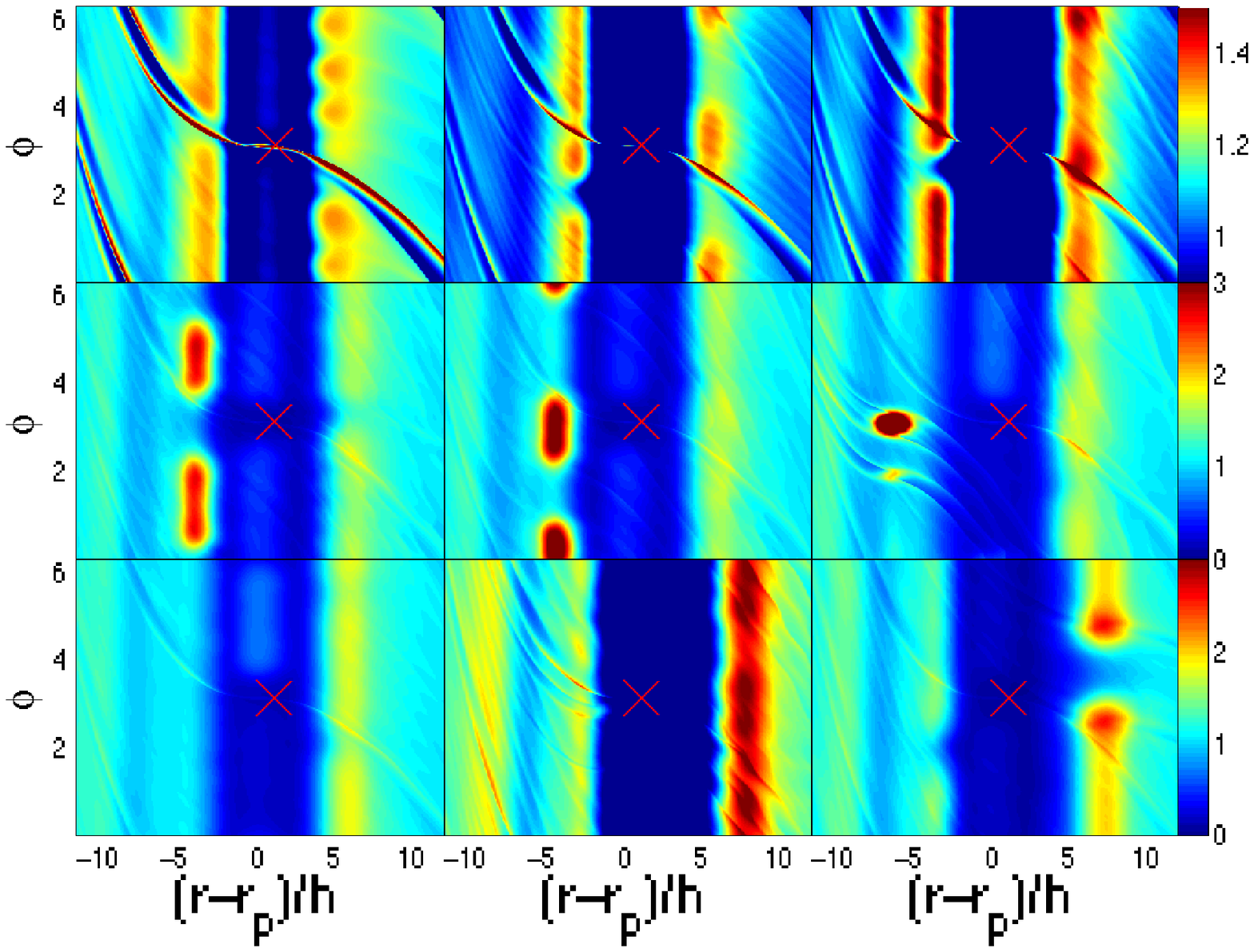,height=6in,width=6in,angle=0}
\end{center}
\caption{Snapshots of the disk surface density
        $\Sigma \cdot r^{3/2}$ at $t = 200$, $600$, $1000$, 
        $4600$, $5000$, $5400$, $6000$, $8000$, and $9180$ orbits 
        from the top left panel to
        the lower right panel, respectively. The viscosity $\nu =
        0$. The  planet location is marked by an ``X''. Vortices are 
         produced as a result of the RVI. These vortices merge,
         migrating inward, and being produced anew by the planet. 
        }
\label{vort}
\end{figure}

For $\nu = 0$, the planet migrates in a more complex way, now strongly
influenced by the RVI.  
Large amplitude oscillations in the semi-major axis evolution appear and 
sometimes exhibit rapid radial drops. Figure \ref{vort} presents several
snapshots of the disk surface density, showing the evolution of RVI.
The vortices exert strong
torques to the planet as they move past the planet. It seems that the
planet's
migration is still inward overall, though it undergoes many oscillations,
reversals, and fast drops (see the black curve in Figure 1). 
Several factors could have contributed to
this type of evolution. First,  the low disk viscosity makes the
shocks stronger (cf. Figure
\ref{reynoldsviscous}), causing a stronger disk response and faster
disk density evolution.  The excitation of RVI is associated with the
inflexion points (which are regions of density depressions) in the
radial profile of potential vorticity (Lovelace et al. 1999; Li et al.
2000). Second, these vortices tend to have slightly different azimuthal speeds
so they will merge (Li et al. 2001; 2005), forming large scale density
structures azimuthally. Third, they are anti-cyclones
with high densities, so they produce their own spiral shocks around
them. Their influence on the surrounding flow and the existence of spiral shocks
lead to an effective angular momentum transport (mostly outward) so
these vortices will gradually migrate inwards (see results also in Li
et al. 2001). This is seen in the current simulations as well. Fourth,
when the vortices migrate away inwardly (on a relatively fast
timescale), the planet migration is subsequently affected because the disk
density profile is significantly changed by these
vortices. Fifth, because the shocks by the planet is 
strong, new generations of vortices are produced after the previous
generation has migrated away. All these highly non-linear effects,
unfortunately, make it very difficult to predict the behavior of
planet migration. 

\begin{figure}
\begin{center}
\epsfig{file=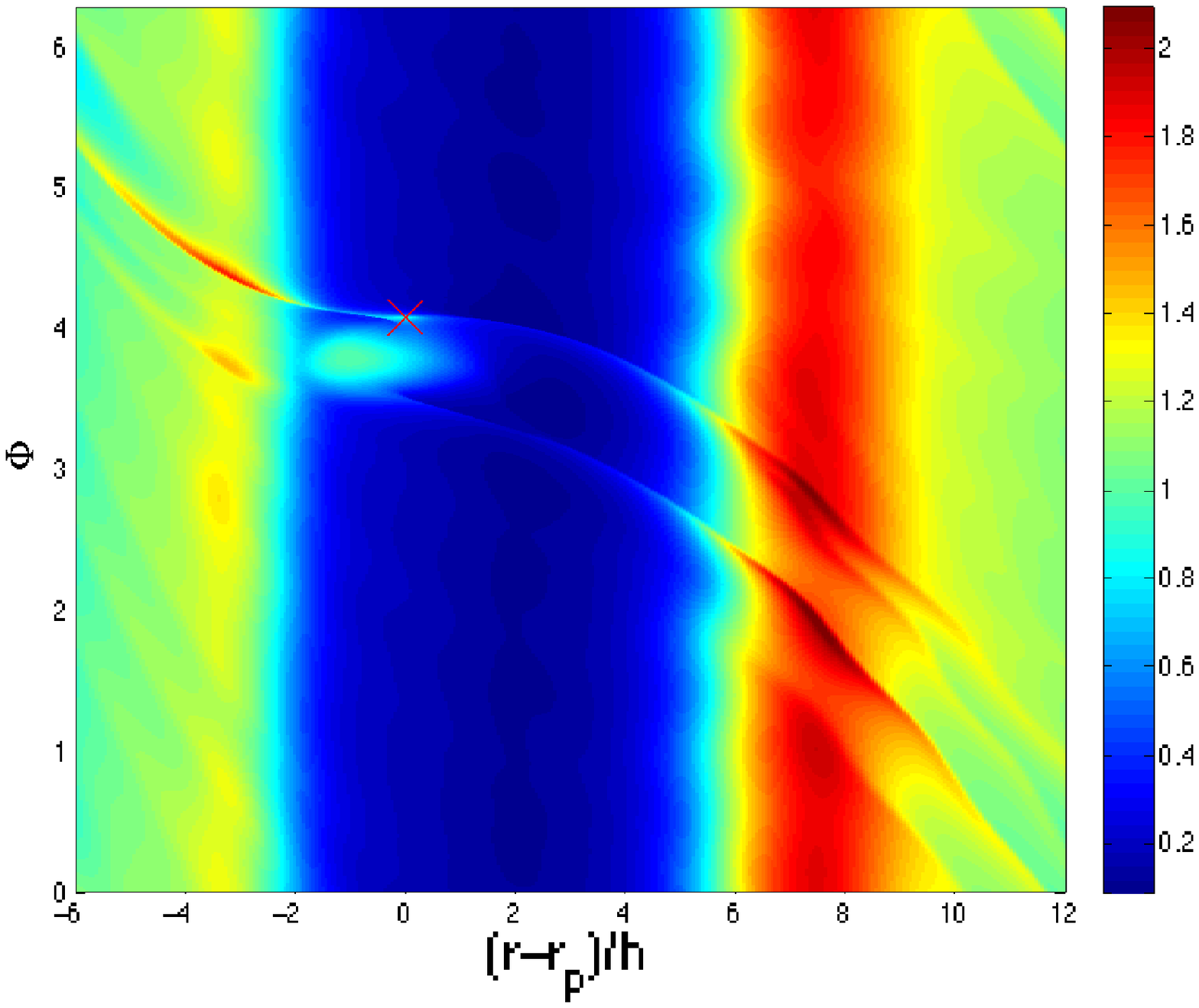,height=5in,width=5in,angle=0}
\end{center}
\caption{The 2-D disk surface density, $\Sigma\cdot r^{3/2}$, is shown.
        A density blob gives rise to a big negative torque at the time of rapid drop 
        (about 7960 orbits).
        }
\label{rho-turn398}
\end{figure}

\begin{figure}
\begin{center}
\epsfig{file=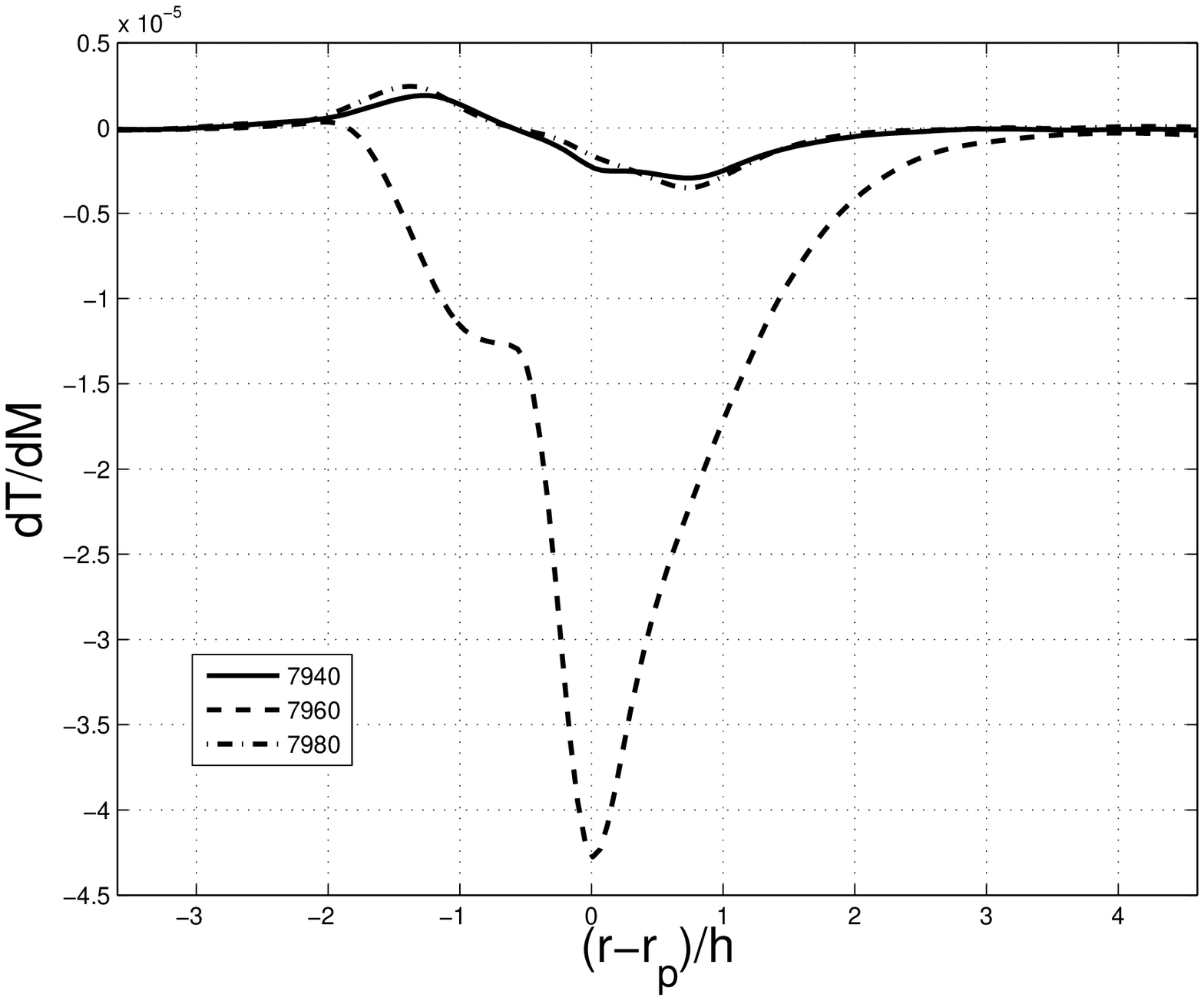,height=3in,width=5in,angle=0}
\end{center}
\caption{
        Torque density distribution before and after the rapid drop of
        the planet ($t \sim 7940 - 7980$ orbits). The large negative
        torque appears when the density blob is located lagging
        behind planet's orbit.
        }
\label{dTdMcmp397-399}
\end{figure}

One curious observation is the fast radial drop during the planet
migration, as indicated, say, between $t = 6400$ and $8000$ orbits
(cf. Figure \ref{09v_runs}). In Figure \ref{rho-turn398}, 
we show the density distribution at the time of a rapid drop (7960
orbits). It is interesting to see that the drop is coincident with
this close encounter between the planet and the density blob.  
The mass of the dense blob is estimated to be the same order
of the planet, about $3\times 10^{-5}$ or $10 M_{\oplus}$. The $dT/dM$
profiles at the time around the rapid drop are shown in Figure
\ref{dTdMcmp397-399}. We can identify that at the time of 7960 orbits, 
a huge negative torque occurs and could contribute to the rapid drop
of the  planet.

\section{Possible 3-D Effects}
\label{sec:3-D}

%

Our results indicate that the disk gas density distribution near the  
planet sensitively controls migration. Migration stoppage in low 
viscosity disks is a  consequence of a systematic mild redistribution 
of gas mass near the planet, favoring  outward over inward
torques. It does not require complete removal of gas near the planet,  
as in the type 2 regime (Li et al 2009). The redistribution is in turn 
controlled by shocks. The location and structure of these shocks have an important  
influence on the feedback torque on the planet. The nature of the shocks
that occur in 3-D can be quite different from the 2-D case analyzed in
this paper. In a 2-D isothermal disk with pressure, only one type of  
wave is excited, a rotationally modified acoustic wave. In a 3-D disk 
that is not  vertically isothermal and/or has a nonzero vertical buoyancy frequency, this wave is  
modified and other types of waves may be excited 
(Lubow \& Pringle 1993; Korycansky \& Pringle  1995).
Their damping properties differ from the 2-D case. If the disk is not
vertically isothermal, as suggested by steady-state models of dead  
zones (e.g., Terquem 2008), then the
main wave that is excited, the $f$ mode, becomes more confined near the  
disk surface as it propagates, through 'wave channelling' (Lubow \& Ogilvie 1998; Bate  
et al 2002). The wave becomes more nonlinear as it propagates and undergoes more rapid shock  
damping than in the 2-D case. Since the material that gets shocked lies above the disk  
midplane, it is not clear how effective the breaking surface waves will be in  
affecting migration in comparison to the isothermal case. But, the rate of change of disk  
angular momentum produced by waves is determined by the angular momentum flux they  
carry. For given disk surface density near a resonance, the $f$ mode carries about the same amount of angular  
momentum flux as the 2-D acoustic mode. So if the $f$ mode damps closer to the planet than  
the 2-D acoustic mode, then its effects on the migration torque could be more important.
It is possible that the upper layers are successively shocked
from the outside-in towards the midplane
and displaced radially. The process may become less effective, as
  the remaining gas becomes less optically thick
and more isothermal. These suggestions are speculative.
Further analysis is required to determine the importance of the $f$ mode  
effects on migration.

Modes other than the $f$ mode that are excited in a 3-D disk can damp  
rapidly. For a
vertically isothermal disk undergoing adiabatic wave perturbations,  
the fraction of the wave
energy that goes into these alternative modes is given by 
$1-\sqrt{\gamma(2-\gamma)}$ [see Eq.  (B4) in Bate et al 2002]. 
For $\gamma = 1.4$, this fraction is only about $8\%$. It is  
possible that the damping of these waves may produce a feedback torque that is more significant  
in strength than $8\%$ of the total feedback torque.  The reason is that the wave damping will  
likely occur closer to the planet than the 2-D mode investigated in this paper. For example,  
vertically propagating gravity waves are produced that damp in the disk atmosphere. 
For a $\gamma =  5/3$ gas, the ratio is $25\%$ and the damping effects of these modes are more important. The  
damping of these waves occurs well above the disk midplane and
it is not clear how much the feedback torque on the planet is  
modified. As discussed above, the disk may be affected from the outside-in, towards the midplane.
A proper 3D analysis of the disk evolution in the low viscosity case  
is required.

\section{Summary and Discussion}
\label{sec:diss}

We have carried out 2-D global hydrodynamic simulations to study the
migration of a $10 M_{\oplus}$ protoplanet in a protoplanetary disk. 
The disk surface density is taken to have the same value in the minimum
mass solar nebula model, but we have taken the normalized disk sound
speed to be relatively low, $c_s = 0.035$. In Paper I, we have
shown the existence and the concrete values for critical planet masses
(depending on the disk mass and sound speed) 
above which the density feedback effects will slow down the type I
migration significantly. Here, we have mainly focused on the long term
behavior of planet migration in such low viscosity disks. We find the
following results:

1) When the disk viscosity is high (e.g., $\nu \geq 10^{-6}$, or $\alpha
\geq 10^{-3}$), the density wave damping is dominated by the disk
viscosity. The migration can be described as the typical type I migration. 

2) When the viscosity is relatively low (e.g., $\nu$ is between $\sim
10^{-8}$ and $10^{-6}$, or $\alpha
\sim 10^{-5}$ and $10^{-3}$), the density wave damping is dominated by
shocks. This then modifies the disk surface density profile quite
significantly, which produces a density feedback effect that alters
the planet migration, slowing it down into a viscous time scale or
halting the migration altogether. The new migration timescale, $t 
\geq 1/\nu \sim 10^6$ orbits, is considerably longer than the usual
type I migration time. This range of the disk $\alpha$ is
interestingly consistent with the expected values in the ``dead
zone'' of protoplanetary disks where protoplanet cores are believed
to arise. If the cores of protoplanets can manage to grow above the
critical masses (as given in Paper I) without migrating away, then
these cores can spend a long time in the dead zone (essentially the
disk lifetime). 

3) When the disk viscosity is even lower (e.g., $\alpha < 10^{-5}$),
the density feedback effect is still present but the RVI starts to
dominate the nonlinear evolution of the disk. The planet migration is
severely affected by the RVI. Large amplitude oscillations appear in the planet
semi-major axis evolution and the rapid drop of the planet occurs sometimes as 
RVI-induced density blob experiences close encounters with the
planet. The overall migration seems still inward and becomes
unpredictable. It is not quite clear whether realistic disks will ever
have such low viscosities. 

We have only studied the long term migration behavior of a $10
M_{\oplus}$ protoplanet. This mass is above the critical mass limit
discussed in Paper I. For lower planet masses, however, even for low
viscosity disks, the shocks produced by the planet will tend to be
weak, so the density feedback effects discussed in this paper and
Paper I will not be strong enough to slow down the migration
significantly. In this limit, the Type I migration still poses a
serious threat to the survivability of these small mass protoplanets  
(say, $< 3 M_{\oplus}$), if no other mechanisms can stop the migration.

The critical masses for stopping planet migration are sensitive to the disk 
interior temperature. Dead zones may have higher temperatures than assumed here. 
For a steady state disk, the surface density varies inversely with $\nu$. 
The higher surface density, due to the lower $\nu$ in a dead zone, gives rise to 
a higher optical depth and therefore higher temperature at
the disk midplane (e.g., Terquem 2008). If the disk temperatures reach a 
value corresponding to $H/r \ga 0.1$, then the critical masses can become substantially 
higher than determined here and the effects of the feedback effect on planet 
migration become much less important.
In addition, realistic 3-D simulations are certainly
desirable to address how layered vertical structures
(with both magnetically active and less-active regions)
will affect the wave damping and planet migration.

\acknowledgments

The research at LANL is supported by a Laboratory Directed Research
and Development program. C.Y. thanks the support from National Natural
Science Foundation of China (NSFC, 10703012) and Western Light
Young Scholar Program. S.L. acknowledges support from NASA Origins
grant NNX07AI72G.

\appendix

\section{Shock Damping for Low Mass Proto-Planets in
  Inviscid Disks}

In this paper, we have studied the density wave damping by
shocks when the disk viscosity is low and how this damping
affects the disk density evolution and planet migration over $10^4$
orbits.  The existence of the
shocks, however, deserves further analysis. Prior studies (e.g.,
Goodman \& Rafikov 2001) have suggested that shocks will always be
produced in an inviscid disk, even for very small planet masses. Other
nonlinear studies (e.g., Korycansky \& Papaloizou 1996) have examined
similar issues, pointing out the importance of parameter
${\it M} = r_H/h = (\mu/3)^{1/3}/h$, where $\mu$ is ratio of planet mass to the central
star. (Their definition omitted the factor of 3.) For $c_s=0.035 -
0.05$, the planet masses considered in Korycansky \& Papaloizou (1996)
are all larger than $10 M_{\oplus}$, which is above the critical
mass limit shown in Li et al. (2009). This means that shocks should
always be produced in an inviscid disk for this planet mass. 
Here, we present some numerical results using 2-D global
hydrodynamic simulations, extending the planet mass to values smaller
than one Earth mass. To isolate the effects of shock production, we
place the planet on a fixed circular orbit and turn off the disk self-gravity. 

\begin{figure}
\begin{center}
\epsfig{file=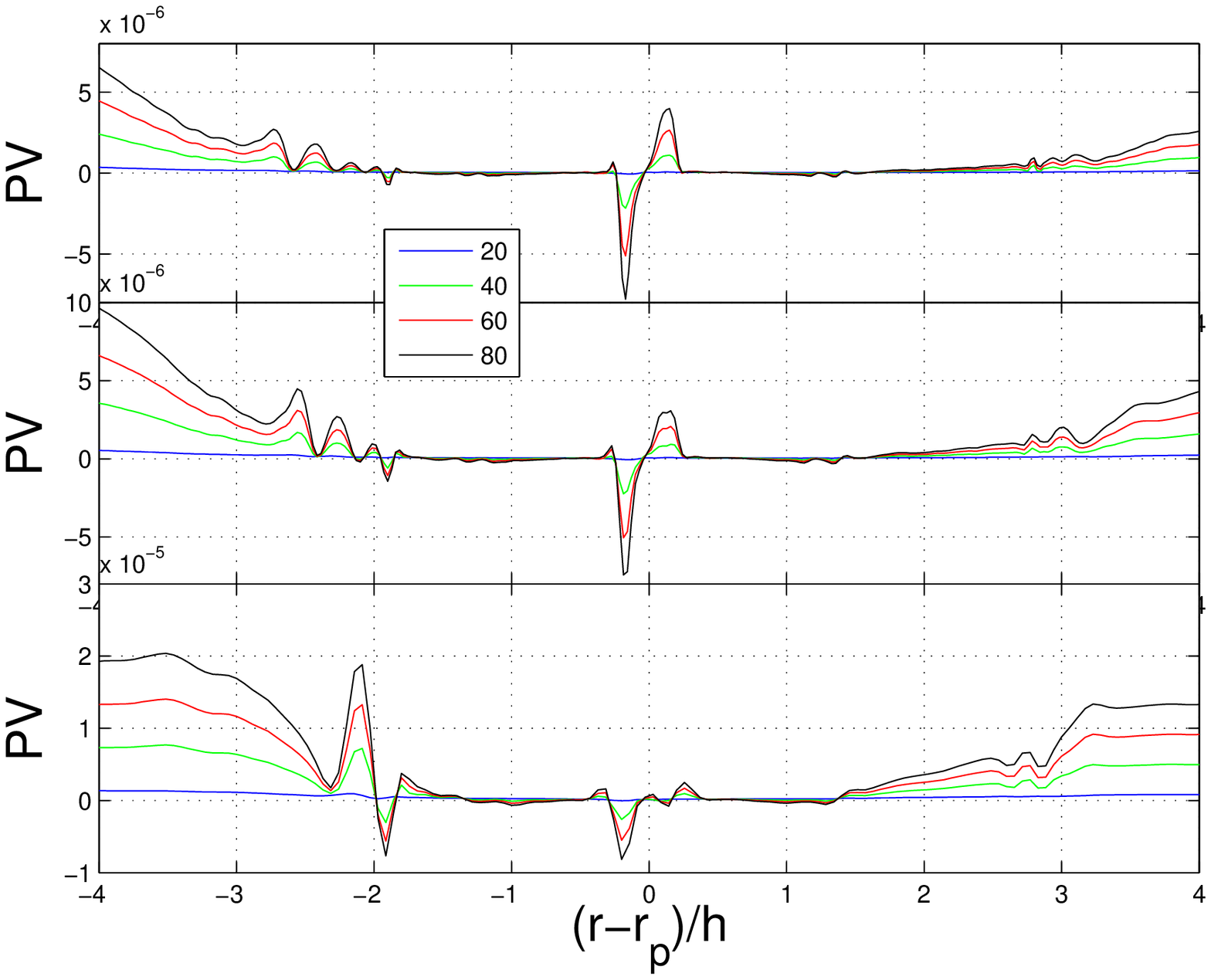,height=3in,width=5in,angle=0}
\end{center}
\caption{The evolution of the azimuthally averaged PV profile at $t =
  20, 40, 60$, and $80$ orbits. The initial PV profile is subtracted
  away in each curve. The planet mass is $0.5 M_{\oplus}$ and the
  disk scale height is $h=0.035$. As time increases, the PV profiles
  deviate more from the initial profile. The resolution of the
  simulations is $n_r\times n_{\phi} = 2000\times 8000, 1600\times 6400$, and $800\times
  3200$, from top to bottom panel, respectively.        
        }
\label{shockresolution}
\end{figure}

Such simulations are numerically challenging because the simulations
have to capture and resolve very weak shocks. We have used an initial
disk surface density profile that goes as $r^{-3/2}$ and the disk is
isothermal and has a constant $c_s$ so that the initial disk potential 
vorticity (PV) profile is
nearly flat (the weak radial dependence of PV from the pressure gradient is
fully captured). We then monitor the changes in PV. Figure
\ref{shockresolution} shows the azimuthally averaged radial PV profile
around a protoplanet of $0.5 M_{\oplus}$. The normalized disk scale height is
$h=0.035$. This gives ${\it M} = r_H/h = 0.23$. We used a constant softening
distance of $\epsilon = 1.0 h$.

In an ideal flow, PV should be conserved.  There are (at
least) two  non-ideal regions in
our simulations. One is the co-orbital region where the planet mass is
introduced. This process is necessarily non-ideal, no matter how
slowly the planet mass is introduced. This will produce changes in
PV. This is because, when the planet is introduced, the flow lines
around the planet change from being non-interacting to horseshoe orbits.
This non-ideal process could produce changes in PV.
Judging from Figure \ref{shockresolution}, this process affects
the PV in a region spreading over $\sim \pm 0.4 h$. The other region
is associated with the density wave propagation where the waves could
steepen into weak shocks (e.g., Goodman \& Rafikov 2001). We believe
this is represented by the PV changes at $r \geq 3.2 h$ and $r \leq -
2.7 h$ using the highest resolution run. 
It is difficult to determine the exact values of the starting
locations of the shocks. Based on Figure \ref{shockresolution}, we
identify $r = 3-3.2 h$ and $r=-(2.6-2.7) h$ as the shock starting
positions at the right and left side of the planet, respectively. 
Note that on both sides of the planet, there exists an ``ideal'' flow
region where the PV remains largely unchanged ($\sim 0.4 - 2.0 h$). 
Even with these high resolutions, the simulations have not
completely converged, though the shock locations are roughly
consistent among the two highest resolution runs. (Note that for
higher planet masses, such as $10 M_{\oplus}$, the shocks are much
stronger and the convergence can be achieved with these resolutions.)

\begin{figure}
\begin{center}
\epsfig{file=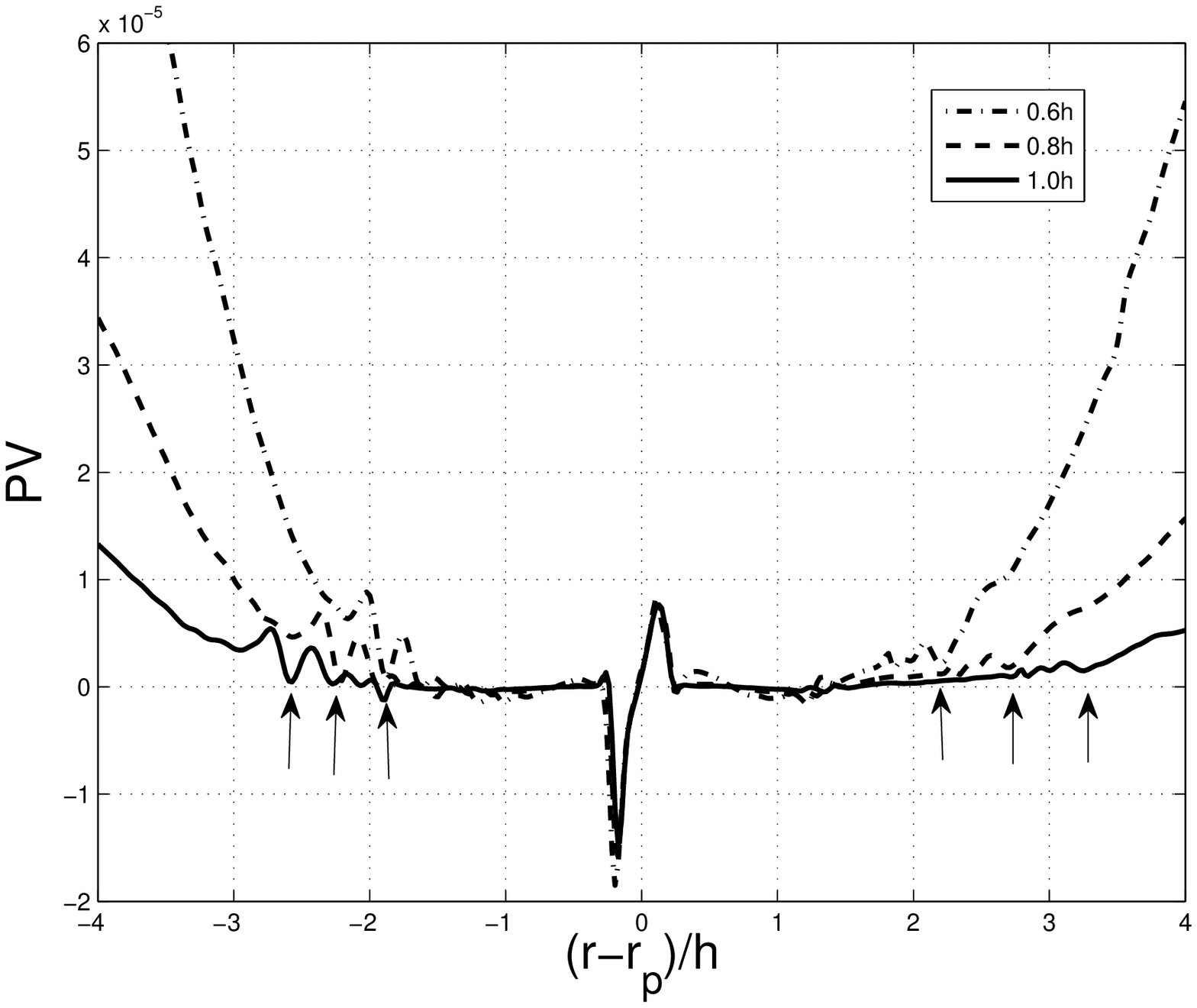,height=3in,width=5in,angle=0}
\end{center}
\caption{
The PV profile at 146 orbits for three different softening distances. 
The resolution is $n_r\times n_{\phi} = 2000\times 8000$ and the planet
mass is $0.5 M_{\oplus}$.
The shock locations are indicated by arrows. 
        }
\label{shocksoftening}
\end{figure}

It is interesting to see that even for ${\it M} = r_H/h$ as low as
$0.23$, shocks are clearly produced. We have also found that the shock
location and strength depend on the softening distance we use. This is
not surprising because larger softening distance will weaken the
strength of wave excitation. Since we do not really know what is the
most appropriate softening to use in a 2-D simulation, we have tried
three values, $\epsilon = 0.6, 0.8$ and $1.0 h$. Figure
\ref{shocksoftening} shows the PV profile produced by a $0.5
M_{\oplus}$ planet in a disk with $h=0.035$. The time is 146 
orbits. The shock gets stronger when the softening is smaller, as
indicated by the magnitude in PV. In addition, the starting location
of the shock moves further away from the planet when the softening
distance increases, as indicated by
the PV profiles to the right of the planet. 

\begin{figure}
\begin{center}
\epsfig{file=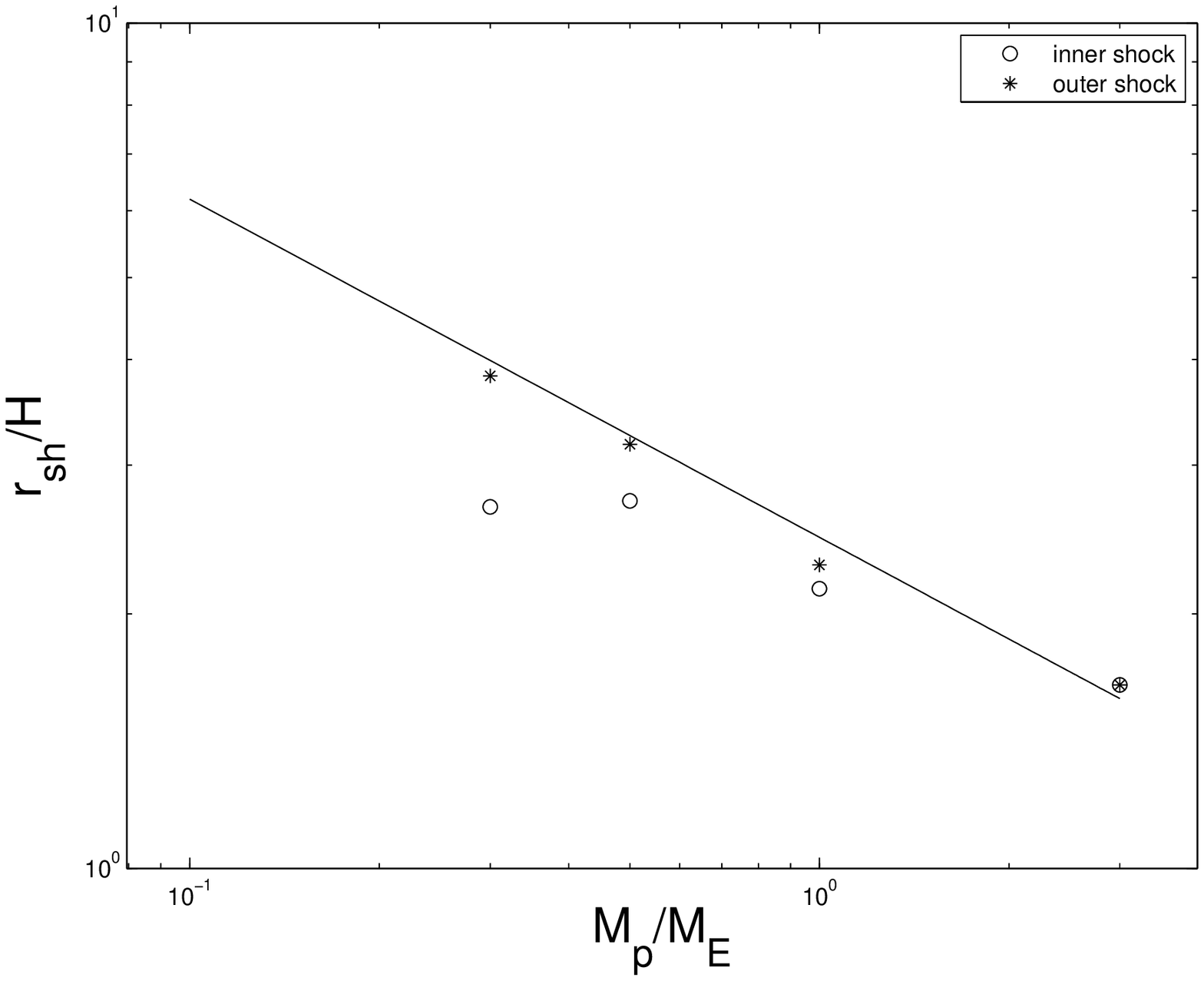,height=3in,width=5in,angle=0}
\end{center}
\caption{The shock locations as functions of planet masses (all 
from $n_r\times n_{\phi}= 2000\times8000$
resolution runs and $1.0h$ softening). 
The solid line is Goodman \& Rafikov's result.       
        }
\label{shocklocation}
\end{figure}

We have also tried to make quantitative comparison between our
simulations and the results by Goodman \& Rafikov (2001). Figure
\ref{shocklocation} shows our best estimates from simulations along
with their predictions [eq. (30)] in Goodman \& Rafikov (2001). 
(Note that their study was done in a shearing sheet configuration.) 
For these simulations we have used
a softening of $1.0 h$. 
The agreement is quite amazing, at least for this choice of softening.
In their analysis, they matched a linear wave
excitation process with the nonlinear propagation. It is hard for us
to ensure that the wave excitation in simulations is strictly in the
linear regime (though increasing softening distance seems to be going
in that direction). 
We have tried to extend our simulations to $0.1 M_{\oplus}$ planet but
we are less confident about the numerical effects, so we omit that 
result.

\end{document}